\documentclass{PoS}
\usepackage{graphicx}
\usepackage{epsfig}
\usepackage{subfigure}
\title{AGILE detection of variable gamma-ray activity from the blazar
  S5~0716+714 during September--October 2007}

\ShortTitle{AGILE detection of variable gamma-ray activity from
  S5~0716+714}

\author{\speaker{F. D'Ammando}$^{ab}$, {A. W. Chen}$^{c}$, {M.
    Villata}$^{d}$, {C. M. Raiteri}$^{d}$, {V. Vittorini}$^{a}$,
    {A. Bulgarelli}$^{e}$, {I. Donnarumma}$^{a}$, {A. Giuliani}$^{c}$,
    {F. Longo}$^{f}$, {L. Pacciani}$^{a}$, {G. Pucella}$^{a}$, {M. Tavani}$^{a}$, {S. Vercellone}$^{c}$ \\
\llap{$^a$}INAF/IASF-Roma, Via del Fosso del Cavaliere 100, I-00133 Roma, Italy\\
\llap{$^b$}Dip. Fisica, Univ. Tor Vergata, Via della Ricerca Scientifica 1,
    I-00133 Roma, Italy\\
\llap{$^c$}INAF/IASF-Milano, Via E. Bassini 15, I-20133 Milano,
    Italy\\
\llap{$^d$}INAF, Osservatorio Astronomico di Torino, Italy\\
\llap{$^e$}INAF/IASF-Bologna, Via Gobetti 101, I-40129 Bologna, Italy\\
\llap{$^f$}Dip. Fisica and INFN Trieste, Via Valerio 2, I-34127 Trieste, Italy\\\\
E-mail: \email{filippo.dammando@iasf-roma.inaf.it}}

\abstract{We report the $\gamma$-ray activity from the Intermediate BL Lac S5
  0716+714 during 2007 September--October observations by the AGILE satellite,
  coincident with a period of intense optical activity of the source monitored
  by GASP--WEBT. 

AGILE observed the source with its two co-aligned imagers, the Gamma-Ray
Imaging Detector (GRID) and the hard X-ray imager (Super-AGILE) sensitive in
the energy range
30~MeV--50~GeV and 18--60 keV respectively, in two different periods: the
  first between 4 and 23 September 2007, the second between 24 October and
  1 November 2007. 

Over the period 7--12 September, AGILE detected $\gamma$-ray
  emission from the source at a significance level of 9.6-$\sigma$ with an average flux
  (E$>$100~MeV) of $(97 \pm 15) \times 10^{-8}$ photons cm$^{-2}$ s$^{-1}$,
  increasing by a factor of at least four within three days. No emission was
  detected by Super-AGILE in the energy range 18--60~keV, with a 3-$\sigma$
  upper limit of 10 mCrab in 335 ksec. The $\gamma$-ray flux of S5 0716+714 detected
  by AGILE is the highest ever detected for this blazar and one of the most
  intense $\gamma$-ray fluxes detected from a BL Lac object. The Spectral Energy
  Distribution (SED) of mid-September seems to be consistent with the synchrotron
  self-Compton (SSC) emission model, but only by including two SSC components
  with different variability.

In October 2007 AGILE repointed toward S5 0716+714 following an intense optical flare, measuring an average flux of
 $(47 \pm 11) \times 10^{-8}$ photons cm$^{-2}$ s$^{-1}$ at a significance
 level of 6.0-$\sigma$. 

The $\gamma$-ray flux during both AGILE pointings appears to be highly variable
 on timescales of 1 day.}

\FullConference{Workshop on Blazar Variability across the Electromagnetic Spectrum\\
		 April 22-25 2008\\
		 Palaiseau, France}

\begin{document}

\section{Introduction}

The source S5 0716+714 was classified by Biermann (1981) as a BL Lac object,
because of its featureless optical spectrum and high linear polarization. The
optical continuum is so featureless that every attempt to determine the
spectroscopic redshift of the source has failed; however, very recently through optical
imaging of the underlying galaxy was estimated a redshift of z = 0.31 $\pm$
0.08 (Nilsson et al. 2008).

The source belongs to the Intermediate BL Lac class according to its
Spectral Energy Distribution. In fact, observations by BeppoSAX (Tagliaferri et
al. 2003) and XMM-$Newton$ (Foschini et al. 2006, Ferrero et
al. 2007) provide evidence for a concave X-ray spectrum in the
0.1--10~keV band, a signature of
the presence of both the steep tail of the synchrotron emission and the flat
part of the Inverse Compton spectrum. The detection in the X-ray band of
fast variability only
in the soft X-ray component can be interpreted as the contemporary presence
of a slowly
variable Compton component and a fast and erratic variable tail of the
synchrotron component.

In general, the variability of this blazar is strong in every band on both
long and short intraday timescales. The optical and radio
historical behaviour has been analyzed by Raiteri et al. (2003), while the
EGRET telescope onboard $CGRO$ (Hartman et
al. 1999) detected S5 0716+714 several times in the $\gamma$-rays (Lin et al. 1995). The integrated flux above 100 MeV varied between (13 $\pm$ 5) and
(53 $\pm$ 13) x 10$^{-8}$ photons cm$^{-2}$ s$^{-1}$.

We present the analysis of the AGILE data
obtained during the S5 0716+714 observations in September--October 2007, in
particular two flaring episodes: the first in mid-September,
the other on 22--23 October 2007. Preliminary results were communicated in
Giuliani et al. (2007) and a more detailed analysis is presented in Chen et
al. (2008).

The strong
$\gamma$-ray flare detected by AGILE in mid-September triggered observations by the
GLAST-AGILE Support Program
(GASP) of the WEBT\footnote[1]{\texttt{http://www.oato.inaf.it/blazars/webt/; see e.g. Villata et al. (2007)}}
(see Carosati et
  al., 2007). About one month later the GASP observed a new very bright phase
of the source, triggering Swift as well as new AGILE observations. In the
period from September to October 2007, S5 0716+714 showed intense activity
with strong optical flaring episodes and a rare contemporaneous optical-radio
outburst (Villata et al. 2008). 

The results of a multiwavelength campaign on S5 0716+714 with
simultaneous AGILE and Swift observations in October 2007 are discussed in
Giommi et al. (2008).
Throughout this paper the quoted uncertainties are given at the
1--$\sigma$ level, unless otherwise stated.

\section{AGILE Observations and Data Analysis} 

The AGILE scientific Instrument (Tavani et al. 2008)
is very compact and combines four
active detectors yielding broad-band coverage from hard X-rays
to $\gamma$-rays: a Silicon Tracker optimized for $\gamma$-ray imaging in the 30
MeV--50 GeV energy band (Prest et al. 2003), a
co-aligned coded-mask X-ray imager sensitive in the 18--60 keV energy band (Feroci et al. 2007), a
non-imaging Cesium Iodide Mini-Calorimeter sensitive in the 0.35--100 MeV
energy band (Labanti et al. 2006) and
a segmented Anticoincidence System (Perotti et al. 2006). 

In September, the AGILE satellite was performing its Science
Verification Phase and devoted three weeks to the observation of S5 0716+714
between 2007 September 4 14:58 UT and September 23 11:50 UT, for a total
pointing duration of $\sim 16.9$ days$\footnote[2]{Between 2007  September 15
  12:52 UT and September 16 12:42 UT AGILE performed a calibration test on the
  Crab pulsar and for two days S5 0716+714 was
  out of the Field of View of the satellite.}$. 

In October, AGILE repointed toward the source and observed S5 0716+714 between
2007 October 24 9:47 UT and November 1 12:00 UT, for a total pointing
duration of $\sim$ 8.1 days.



Level--1 AGILE-GRID data were analyzed using the
AGILE Standard Analysis Pipeline as described in detail in Vercellone et
al. (2008). Counts, exposure and Galactic background $\gamma$-ray maps are
created with a bin size of $0.\!\!^{\circ}3 \times 0.\!\!^{\circ}3$ for E $>$
100 MeV. To reduce the particle background contamination we selected only
events flagged as confirmed $\gamma$-ray events (\emph{filtercode=5}) and all events
collected during the South Atlantic Anomaly (SAA) were rejected
(\emph{phasecode=18}). We also rejected all the $\gamma$-ray events whose
reconstructed directions form angles with the satellite-Earth vector smaller
than 80$^{\circ}$ (\emph{albrad=80}), reducing the $\gamma$-ray Earth
Albedo contamination by excluding regions within $\sim$ 10$^{\circ}$ from the
Earth limb.

The average $\gamma$-ray flux as well as the daily values were derived according to Mattox et al. (1993).
First, the entire period was analyzed to determine the diffuse emission parameters; then, the source flux density was estimated
independently for each of the 1-day periods with the diffuse
parameters fixed at the values obtained in the first step.



 \section{Results} 

%
S5 0716+714 was detected by the GRID instrument onboard AGILE in the period 7--12 September 2007, with the source at about 15$^{\circ}$
off-axis, at a significance level of 9.6-$\sigma$ with an average $\gamma$-ray
flux of $ F_{E > 100~\rm MeV}  = (97 \pm 15) \times 10^{-8}$\,photons cm$^{-2}$
s$^{-1}$, as derived from a maximum likelihood analysis. The peak level of the
$\gamma$-ray flux is $ F_{E > 100~\rm MeV}  = (193 \pm 42) \times
10^{-8}$\,photons cm$^{-2}$ s$^{-1}$, showing an increase of the flux by a
factor four within three days; this flux is the highest ever detected from
S5 0716+714. 

Super-AGILE observed S5 0716+714 between 7 and 12 September 2007
for a total on-source net exposure time of 335~ksec and the source was not
detected above 5-$\sigma$ by the Super-AGILE Iterative Removal Of Sources
(IROS) applied to the image, in the 20--60~keV energy range. A 3-$\sigma$
upper limit of 10 mCrab was obtained from the observed count rate by a study
of the background fluctuations at the position of the source and a simulation
of the source and background contributions with IROS.

%

%
%

\begin{figure}[!th] 
    \centering
    \includegraphics[width=.50\textwidth]{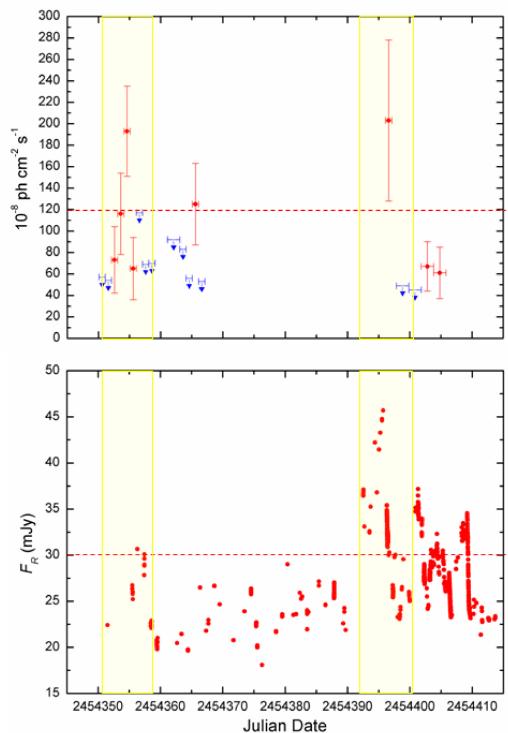}
    \caption{
      In the top panel, AGILE--GRID $\gamma$-ray light curve with
      1-day or 2-day resolution for fluxes in units of
      $10^{-8}$\,photons cm$^{-2}$ s$^{-1}$  for $E > 100$~MeV . The downward arrows represent
      2-$\sigma$ upper-limits. In the bottom panel, the $R$-band optical light
      curve as observed from GASP-WEBT. The mean flux density level is
      highlighted with horizontal red dashed line. The yellow shaded regions
      indicate the two high activity periods of the source in the $\gamma$-ray band.}
    \label{figure1}
\end{figure}

A comparison between the $\gamma$-ray and optical light curves is shown in
Figure~\ref{figure1}: the top panel shows the $\gamma$-ray light curve
with 1 or 2 day resolution for photons above 100~MeV, the bottom panel shows
the $R$-band optical light curve as obtained by the GASP-WEBT. The results of
the GASP-WEBT multifrequency monitoring of 0716+714 in September--October 2007 are presented in Villata et al. (2008). During the September observation the $\gamma$-ray flux of the source appears to
be highly variable on timescales of 1 day.

Moreover, S5 0716+714 in mid-October shows increasing optical flux, reaching a peak
of $F_R$ = 45.7 mJy on October 22.2 (Villata et al. 2008); at that time S5 0716+714,
even if rather off-axis ($\sim$ 50$^\circ$ from the axis) was seen by AGILE to
have a high $\gamma$-ray flux. In particular, between
2007 October 22 12:33 UT and October 23
12:06 UT the maximum likelihood analysis provides a significance of 4.0-$\sigma$
with a flux of $ F_{E > 100 \rm MeV}  = (203 \pm 75) \times
10^{-8}$\,photons cm$^{-2}$ s$^{-1}$.  Note, however, that AGILE has a high particle
background at high off-axis angles, and that the exposure time is relatively
short.

After this flaring episode AGILE observed the source with a dedicated
repointing during the period between October 24 9:47 UT and November 1
12:00 UT, and over the entire period detected a $\gamma$-ray flux above 100
MeV at a significance level of 6.0-$\sigma$ with a lower average
flux of
$ F_{E > 100 \rm MeV}  = (47 \pm 11) \times 10^{-8}$\, photons cm$^{-2}$
s$^{-1}$.

During the September--October observations AGILE detected S5 0716+714 at two
different levels of activity. The $\gamma$-ray spectrum during the high activity state of mid-September
can be fitted with a power law with a photon index of $\Gamma$ = 1.56
$\pm$ 0.30, while during the October ToO the source was in a low
$\gamma$-ray activity state and the photon index of the differential energy spectrum is $\Gamma$ = 1.95
$\pm$ 0.54 (see Fig. 2, left panel). The photon index was obtained with the weighted least
squares method considering only 3 energy bins: 100--200 MeV, 200--400 MeV
and 400--1000 MeV.

\begin{figure}[!t]
\centering
\subfigure
{\includegraphics[width=.462\textwidth]{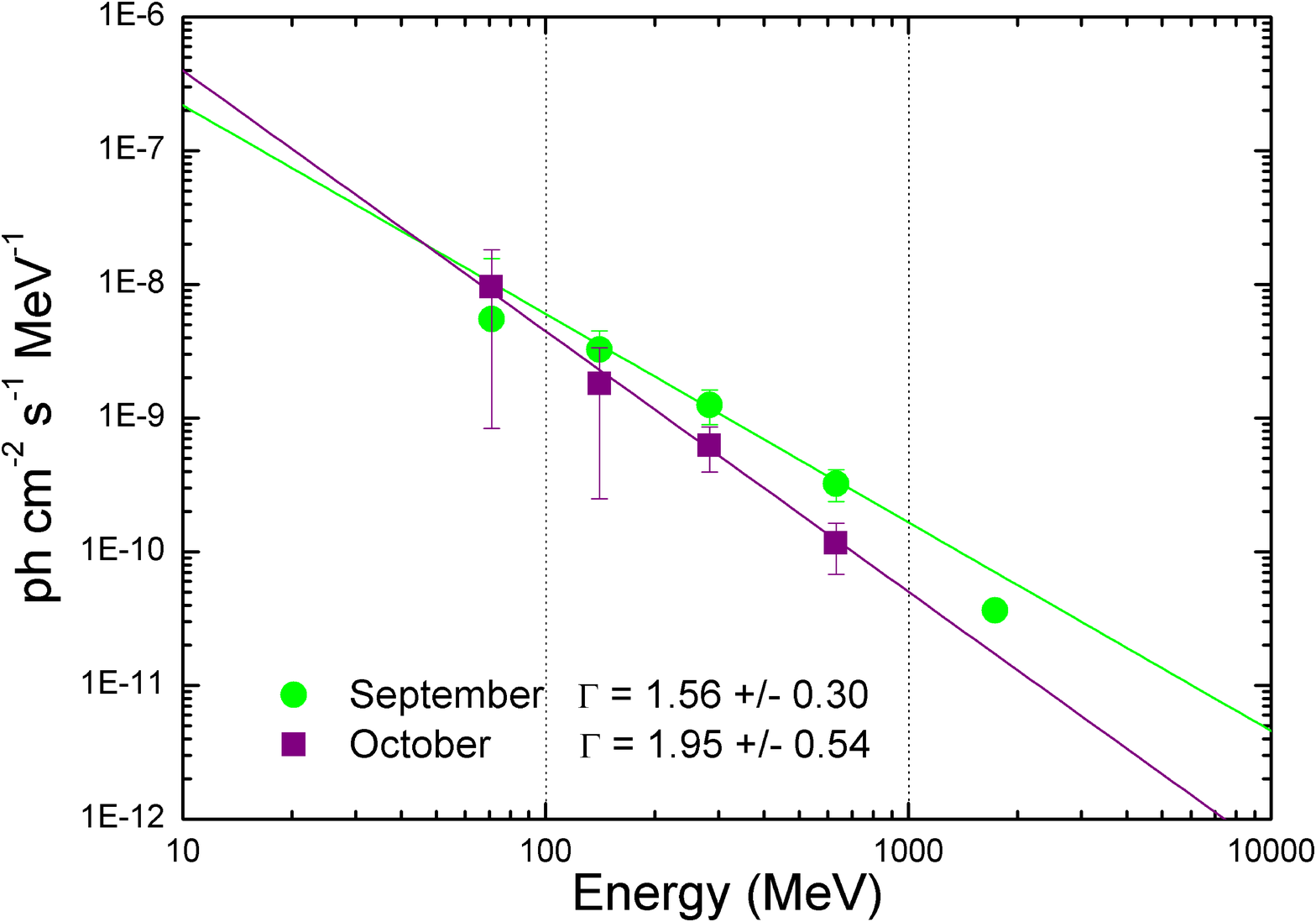}}
\hspace{2mm}
\subfigure
    {\includegraphics[width=.458\textwidth]{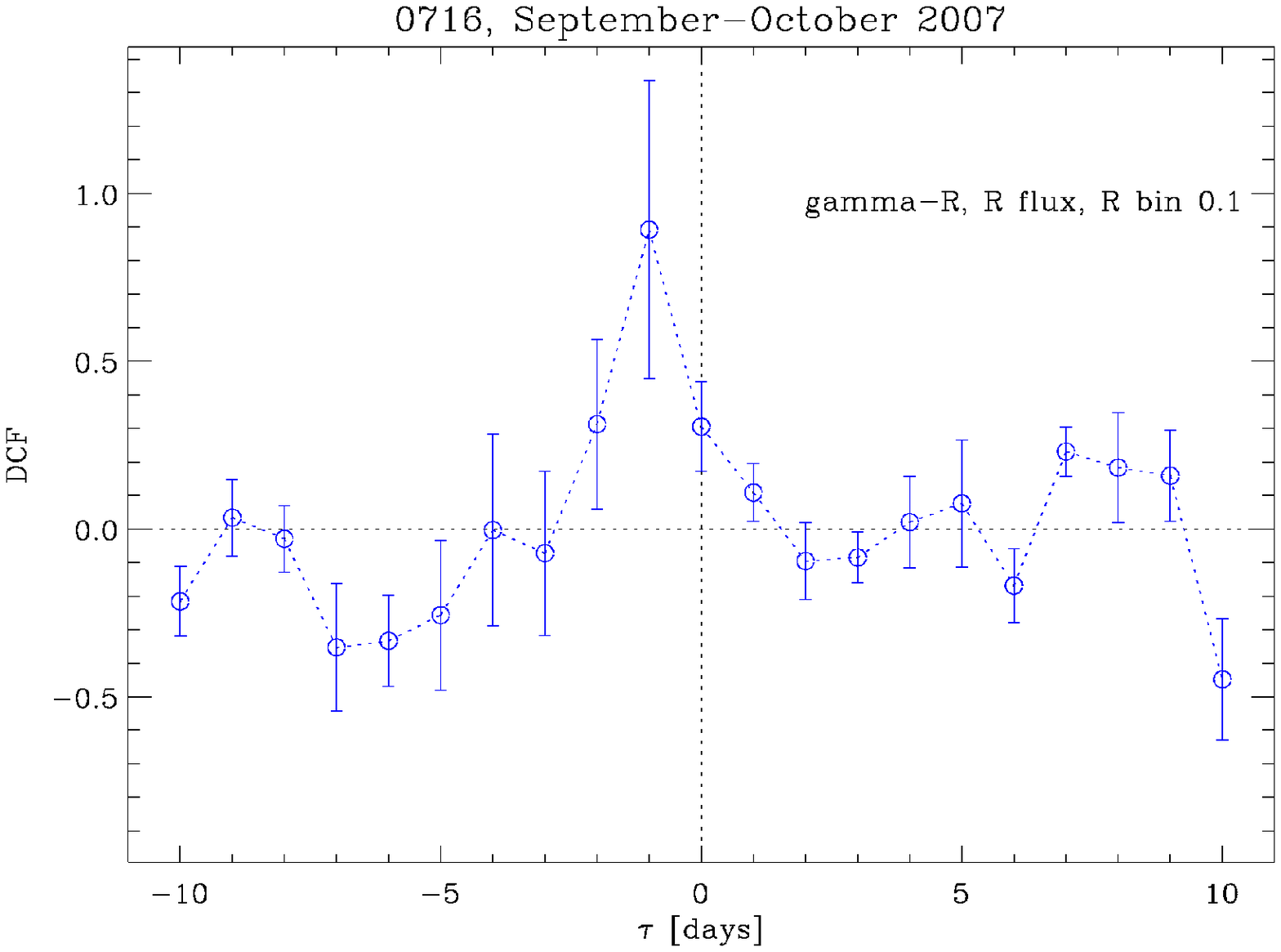}}
 \caption{ \textsl{Left panel:} Gamma-ray photon spectrum
      of S5~0716+714 during the high state of mid-September 2007 (green line)
      and the low state of end October 2007 (purple line). \textsl{Right panel:}
      Discrete correlation function (DCF) between the $\gamma$-ray and $R$-band light curves for S5 0716+714 in September-October 2007.}
\end{figure}

\section{Discussion} 


To analyze the gamma-optical correlation we applied the Discrete Correlation Function
(DCF; see Edelson $\&$ Krolick 1988; Hufnagel and Bregman 1992; Peterson 2001)
to the $\gamma$-ray and $R$-band light curves. The $R$-band flux densities
were averaged over 0.1 day bins to smooth the intranight variability.
The DCF is a statistical method that was developed to analyze unevenly sampled
data trains. The DCF displays a significant peak (DCF $\sim$ 0.9)
at a lag of -1 day (Fig. 2, right panel). Despite the large uncertainty due to poor $\gamma$-ray
sampling, this result suggests a possible delay of the $\gamma$-ray flux variations with respect to the optical ones on the order of 1 day.
Looking at Fig. 1, one can see that most of the DCF signal comes from the quasi-simultaneity of the
$\gamma$-ray and optical peaks of late October (JD $\sim$ 2454396-397). 


We notice that when the $\gamma$-ray fluxes are $\leq$ 120 $\times$
10$^{-8}$\, photons cm$^{-2}$ s$^{-1}$, the corresponding optical
flux densities are around 25--30 mJy. In contrast, the October
$\gamma$-ray peak reaching $\sim$ 200 $\times$ 10$^{-8}$\, photons
cm$^{-2}$ s$^{-1}$ has an optical counterpart of 40--45 mJy (see Fig. 1). This
suggests that a strong optical event simultaneous to the $\gamma$-ray
flare was missed in September, since it occured at the beginning of the
optical observing season as well as the start of the GASP activity. 

The gamma variability seems to depend on the optical flux density changes
roughly quadratically and this would favour a SSC interpretation, in which
the emission at the synchrotron and IC peaks is produced
by the same electron population, which self-scatters the
synchrotron photons. In this case, the 1-day time lag in the high-frequency
peak emission found from the DCF could be due to the light travel time of the
synchrotron seed photons which scatter the energetic electrons.

\begin{figure}[!t] 
\centering
  \includegraphics[angle=0,scale=0.083]{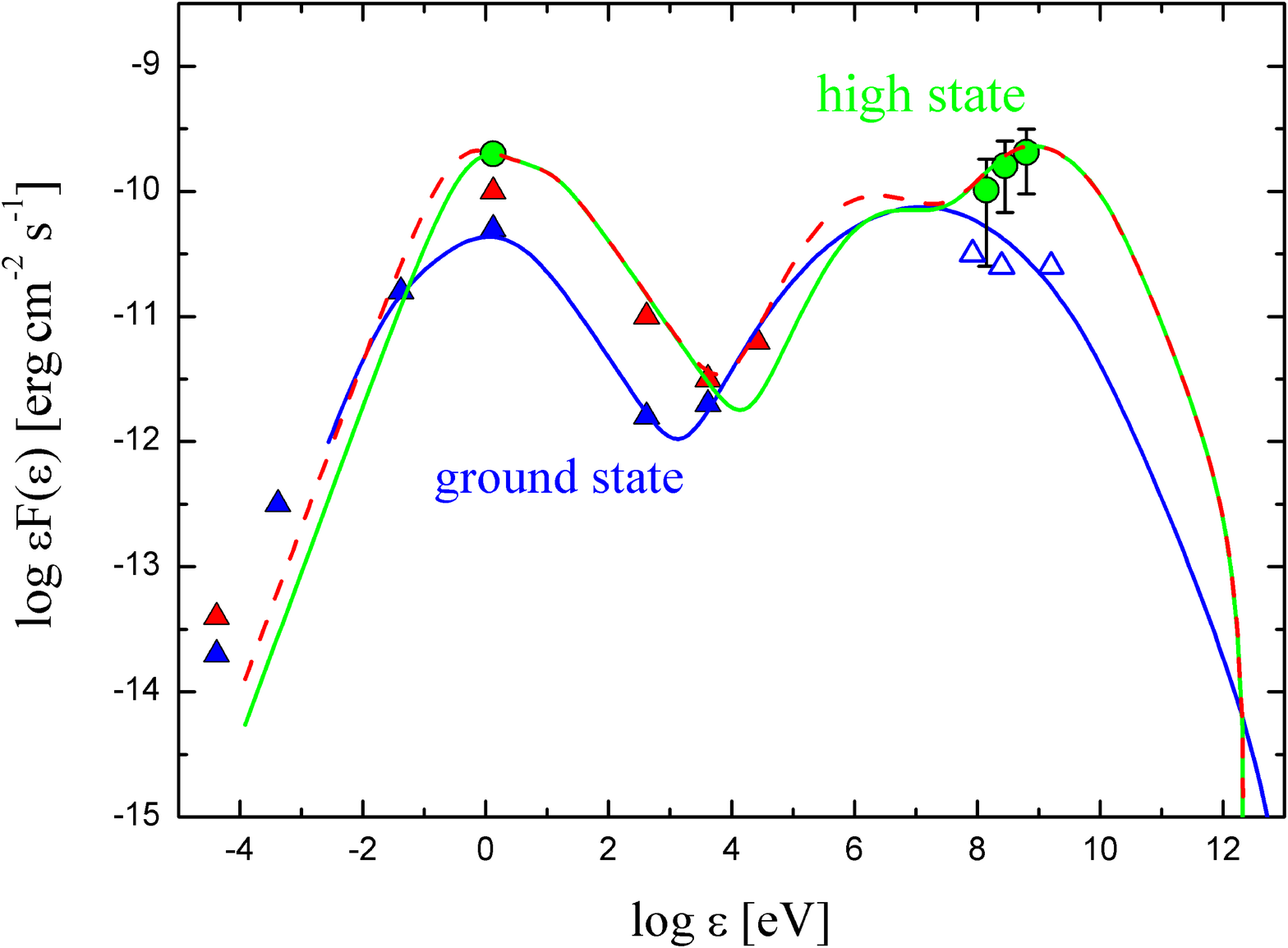}
     \caption{
       The SED of S5 0716+714 including GASP-WEBT optical data
       quasi-simultaneous with AGILE-GRID $\gamma$-ray
       observation in September 2007 (green dots). Historical data over the entire
       electromagnetic spectrum relative to a ground state of the source together
       with EGRET non-simultaneous data is
       represented with blue triangles. Red triangles represent historical data simultaneous with a high X-ray state.}
     \label{figure3}
\end{figure}

The Spectral Energy Distribution with the AGILE and GASP-WEBT data
of September 2007 is shown in Figure~\ref{figure3} as green dots. The
blue solid line shows a simple SSC model fitting simultaneous observations
of a ground state (see Tagliaferri et al. 2003
and references therein) together with non-simultaneous EGRET data (empty blue triangles).
Because the high state of mid-September 2007 cannot be fitted by a
one-zone SSC component alone, we used a model with two SSC components.
Without simultaneous X-ray data the spectrum is poorly
constrained, then we show two models: one with a high hard X-ray state (red dashed line)
and one with a low hard X-ray state (green solid line).

The first SSC component dominates in the optical and X-ray bands and it is
reproduced  with a double power law electron distribution: the spectral index
is $p_{low}=2$ from $\gamma_{min}$ to $\gamma_{break}$ and $p_{high}=4.5$
above $\gamma_{break}$. For the high X-ray state model $\gamma_{min}$ = 500, while
the low X-ray state model has $\gamma_{min}$ = 700; in both cases
$\gamma_{break}$ = 10$^3$. The density at the spectral break is $n_{e}=40\,cm^{-3}$, the blob radius
$R=2 \times 10^{16}$ cm and the magnetic field $B=3$ Gauss.

The second SSC component contributes primarily to the gamma range of the
SED and, to a lesser extent, to the optical emission. The electron
distribution is a single power law with $p=4.5$,
$\gamma_{min}=4 \times 10^{3}$ and $n_{e}=50\,cm^{-3}$. The blob radius is
$R=10^{16}$ cm and the magnetic field $B=1.3$ Gauss. Both blobs are moving with
bulk Lorentz factor $\Gamma=15$, at an angle of $3^{\circ}$ with respect the line of sight.

We cannot exclude a second component due to an external seed
photon field (e.g. mirrored by a putative broad line region) which could also
account for the possible 1-day time lag, but the large
amplitude of $\gamma$-ray variability with respect to that of the optical one
favours a SSC explanation.


\begin{acknowledgments}The AGILE Mission is funded by the Italian Space Agency
  (ASI) with scientific and programmatic participation by the Italian Institute
of Astrophysics (INAF) and the Italian Institute of Nuclear Physics (INFN). We
  wish to express our gratitude to the Carlo Gavazzi Space, Thales Alenia
  Space, Telespazio and ASDC/Dataspazio Teams that implemented the necessary
  procedures to carry out the AGILE re-pointing. The optical data
  presented in this paper are stored in the GASP-WEBT archive; for questions
  regarding their availability, please contact the WEBT President Massimo
  Villata.
\end{acknowledgments}

\end{document}